\documentclass[12pt]{article}

\newcommand{\ra}{\rightarrow} 
\newcommand{\nn}{\nonumber}

\newcommand{\al}{\alpha}

\newcommand{\ov}{\overline}

\newcommand{\be}{\begin{equation}}
\newcommand{\ee}{\end{equation}}
\newcommand{\bea}{\begin{eqnarray}}
\newcommand{\eea}{\end{eqnarray}}
\newcommand{\cI}{{\cal I}}
\newcommand{\FL}{{(-1)^{F_L}}}
\newcommand{\F}{{(-1)^F}}
\newcommand{\kp}{{k_+}}
\newcommand{\km}{{k_-}}
\newcommand{\Dthreep}{{D3+}}
\newcommand{\Dthreem}{{D3-}}

\newsavebox{\orbns}
\newsavebox{\ns}
\newsavebox{\dbrane}
\newsavebox{\dnquiver}
\newsavebox{\orb}
\newsavebox{\bstack}

\title{\vspace{-1in}\parbox{\linewidth}{\small\hfill\shortstack{IASSNS-HEP-98/62}}
\vspace{0.6in}\\
$D_n$ Quivers From Branes}
\author{ Anton Kapustin \thanks{Research supported in part by DOE grant DE-FG02-90-ER40542}\\
{\sl\small School of Natural Sciences, Institute for Advanced Study}\\
{\sl\small Olden Lane, Princeton, NJ 08540}}

\begin{document}
\begin{titlepage}
\renewcommand{\thepage}{ }
\renewcommand{\today}{ }
\thispagestyle{empty}

\maketitle

\begin{abstract}
D-branes can end on orbifold planes if the action of the orbifold group includes
$(-1)^{F_L}$. We consider configurations of D-branes ending on such orbifolds and study the
low-energy theory on their worldvolume. We apply our results to
gauge theories with eight supercharges in three and four dimensions.
We explain how mirror symmetry for $N=4$ $d=3$ gauge theories with gauge group
$Sp(k)$ and matter in the antisymmetric tensor and fundamental representations
follows from S-duality of IIB string theory. We argue that some of these
theories have hidden Fayet-Iliopoulos deformations, not visible classically.
We also study a class of finite $N=2$ $d=4$ theories (so-called $D_n$ quiver theories)
and find their exact solution.
The integrable model corresponding to the exact solution is a Hitchin system on an 
orbifold Riemann surface. We also give a simple derivation of the S-duality group
of these theories based on their relationship to $SO(2n)$ instantons on 
${\bf R}^2\times{\bf T}^2$.

\end{abstract}
\end{titlepage}

\section{Introduction}
Constructing field 
theories from brane configurations in string theory can be of great help in understanding the 
properties of the former (see Refs.~\cite{HW,Witten,Witten1} for some examples and 
Ref.~\cite{GK} for a review.) In this letter we use this approach to investigate certain aspects
of gauge theories with eight supercharges in three and four dimensions. One of our results is  
a simple proof of mirror symmetry for $N=4$ $d=3$ 
theories with gauge group $Sp(k)$ and matter in the antisymmetric tensor
and fundamental representations~\cite{IS,Berkeley}. Our argument is based on S-duality
of IIB string theory and is completely
analogous to that for unitary groups~\cite{HW,Berkeley}. (A different derivation of
mirror symmetry covering both unitary and symplectic groups was presented in 
Ref.~\cite{mfromM}.)
We also argue that $Sp(k)$ gauge theories
with an antisymmetric tensor and two or three flavors of fundamentals have accidental $U(1)$
symmetries in the infrared and therefore hidden
Fayet-Iliopoulos (FI) deformations. (Classically, $Sp(k)$ gauge theories do not admit FI
deformations.) Another result is an exact solution of a class of finite $N=2$ $d=4$ gauge theories, 
so-called $D_n$ quiver theories~\cite{KMV}.
Recall that quiver gauge theories are finite if and only if the quiver is an affine 
Dynkin diagram of type $A,D,$ or $E$~\cite{KMV}. The exact solutions of these theories 
have been found in Ref.~\cite{KMV} via geometric engineering. The $A_n$ case can also be 
solved using M-theory~\cite{Witten} or compactification to three dimensions~\cite{Kap}.
In this paper we extend the latter approach to $D_n$ quivers. We explain how the exact solution
of $D_n$ quiver theories is related to the moduli space of $D_n$ instantons on
${\bf R}^2\times {\bf T}^2$. We use this correspondence to compute the S-duality group
of the theories. The results are in agreement with the geometric engineering approach.
Further, we show that the solution of the model is encoded in the moduli space of
Hitchin equations on an orbifold Riemann surface and derive the corresponding
Seiberg-Witten curve.

The brane configurations that we use are somewhat unusual: they involve D-branes ending on 
orbifold planes. Let us remind why it is possible for D-branes to end on certain orbifolds~\cite{Sen1}.
Consider an orientifold 5-plane with negative charge and a D5-brane on top of it. Such
a configuration has sixteen unbroken supersymmetries.
It also has a vanishing total Ramond-Ramond charge, and therefore a very 
simple S-dual: an orbifold ${\bf R}^6\times {\bf R}^4/\cI$, where $\cI$ is a product of $\FL$ and 
the inversion of all four coordinates of ${\bf R}^4$~\cite{Sen0}. (The S-dual of
an $O5^-$ plane alone is an object which is magnetically charged with respect
to the Neveu-Schwarz B-field. It does not have a simple conformal field theory description.)
Since fundamental strings and 
D3-branes can end on D5-branes, S-duality of IIB string theory implies that D-strings and D3-branes 
must be able to end on such 
orbifold planes. The boundary states describing D-branes ending on orbifolds have been constructed
in Ref.~\cite{Sen1}. To be more precise, the theory on the orbifold worldvolume is a $(1,1)$ $d=6$
theory with gauge group $SO(2)$. (The $SO(2)$ vector multiplet arises from the twisted closed string 
sector.) The end of a D3-brane is a 2-brane magnetically charged with respect to the $SO(2)$
gauge field.
This charge can be positive or negative, so there are two possible boundary states describing a 
D3-brane ending on the orbifold. We will call them $|\Dthreep\rangle$ and $|\Dthreem\rangle$.
Similarly, the end of a D-string is a 0-brane electrically charged with respect to the $SO(2)$.

What is the analogue of this in IIA string theory? There one starts with an $O4^-$ plane
and asks about its strong coupling limit, i.e. about its lift to M-theory. 
One finds~\cite{DasMu,Witten0} that the corresponding M-theory background is 
${\bf R}^5\times {\bf S}^1\times {\bf R}^5/{\bf Z}_2$, where ${\bf Z}_2$ acts as an inversion 
on ${\bf R}^5$ and in addition flips the sign of the 3-form potential. Now we can go back to
IIA string theory by compactifying one of the directions of ${\bf R}^5/{\bf Z}_2$.
The resulting object is an orbifold 5-plane of some kind which is magnetically charged 
with respect to the B-field. Its charge is minus the charge of the NS5-brane~\cite{DasMu,
Witten0}. It is related to NS5-branes in the same way as an $O4^-$ plane is related 
to D4-branes, so one could call it a ``Neveu-Schwarz orientifold.''
This object does not have a simple CFT description.
However, if we put an NS5-brane on top of it, the total charge becomes zero. The
conformal field theory description of this composite object is very simple:
it is a IIA orbifold ${\bf R}^6\times {\bf R}^4/\cI$~\cite{Kutasov,Sen0}, where ${\cal I}$
is the same as before. It can be
shown that the twisted closed string sector gives rise to a single $(2,0)$ tensor multiplet
propagating on the orbifold plane~\cite{Kutasov}. This tensor multiplet describes 
small fluctuations of the NS5-brane sitting at the orbifold plane.
Furthemore, since the composite orbifold contains an NS5-brane, a D4-brane must be able
to end on it. The end of the D4-brane is a 3-brane magnetically charged with respect
to the periodic scalar which is a part of the tensor multiplet.
This charge can be of either sign.

The organization of this paper is as follows.
In Section 2 we investigate the low-energy theory of D-branes ending on orbifolds. 
In Section 3 we use D3-branes ending on orbifolds to study mirror symmetry in $d=3$. 
In Section 4 derive the Seiberg-Witten solution for $D_n$ quiver theories in $d=4$.

\section{The brane configuration}

Consider an orbifold 5-plane ${\bf R}^6\times {\bf R}^4/\cI$ extending in the 
$x^0,x^1,\ldots,x^5$ directions and localized
at $x^6=\cdots=x^9=0.$ Let $k$ D3-branes extend in the $x^0,x^1,x^2,x^6$ directions and 
end on the orbifold plane. Such a configuration preserves eight supersymmetries. This is
true even if some D3-branes carry opposite $SO(2)$ magnetic charges~\cite{Sen1}.
This can be seen from the partition function for an open string
with ends on a pair of such D3-branes. The partition function can be represented as
\be\label{notach}
\int_0^\infty dl\ \langle \Dthreep| e^{-l H_{cl}} |\Dthreem\rangle,
\ee
where $|\Dthreep\rangle,\ |\Dthreem\rangle $ are boundary states describing D-branes
and $H_{cl}$ is the closed string Hamiltonian.
This expression can be evaluated using the formulas in Ref.~\cite{Sen1}. The matrix
element in Eq.~(\ref{notach}) does not exhibit a divergence for $l\ra 0$, which would be 
a signal of tachyon instability. Thus D3-branes with opposite $SO(2)$ charge do not exert
force on each other. Similarly, there is no net force between D3-branes with the
same $SO(2)$ charge.

If not for the orbifold, the low-energy theory on the world-volume of D3-branes
would be an $N=4$ $U(k)$ gauge theory. The orbifold projection breaks supersymmetry down to $N=2$.
Furthemore, if $\kp$ D3-branes have $SO(2)$ charge $+1$, and $\km$ of them have $SO(2)$ charge
$-1$ ($k=\kp+\km$), the projection breaks $U(k)$ down to $U(\kp)\times U(\km)$. 
This can be seen from the fact that only D3-branes with the same $SO(2)$ charge can be 
regarded as indistinguishable, and therefore the Weyl group is reduced from $S_k$ to 
$S_\kp\times S_\km$. To see the breaking of the gauge group more explicitly, notice that the 
partition function for an open string with both ends on D3-branes of the same $SO(2)$ charge 
is given by~\cite{Sen1}
\be\nn
\int_0^\infty dl\ \langle \Dthreep| e^{-l H_{cl}} |\Dthreep\rangle = \int_0^\infty
\frac{dt}{2t}\ {\rm Tr}\ e^{-2tH_o}\frac{1+\F}{2}\frac{1+g}{2}, 
\ee
where $g$ is the inversion of $x^6,\ldots,x^9$. On the other hand, the partition function
of an open string with ends on D3-branes of opposite $SO(2)$ charge is
\be\nn
\int_0^\infty dl\langle\ \Dthreep| e^{-t H_{cl}} |\Dthreem\rangle = \int_0^\infty
\frac{dt}{2t}\ {\rm Tr}\ e^{-2tH_o} \frac{1+\F}{2}\frac{1-g}{2}.
\ee
In the first case, only the string modes which are invariant with respect to $g$ are retained.
In the second case, they are projected out, and instead only the antiinvariant ones are
retained. Any $U(k)$ transformation which is not in $U(\kp)\times U(\km)$ mixes the two types
of open string states and is inconsistent with the orbifold projection. 

Let us consider in more detail how the orbifold projection acts on the adjoint scalars living on 
the worldvolume of D3-branes.\footnote{The following argument is due to A. Sen.}
Let $\alpha$ run from $3$ to $5$, and let $i$ run 
from $7$ to $9$. The Higgs fields $\phi^\alpha$ describe the oscillations of D3-branes
in the directions parallel to the orbifold plane, while $\phi^i$ describe the oscillations
transverse to it. The discussion in the previous paragraph implies that they satisfy
\bea\label{proj}
\phi^\alpha(x^0,x^1,x^2,-x^6)=\Omega \phi^\alpha(x^0,x^1,x^2,x^6) \Omega^{-1}, \\ \nn
\phi^i(x^0,x^1,x^2,-x^6)=-\Omega \phi^i(x^0,x^1,x^2,x^6) \Omega^{-1},
\eea
where $\Omega$ is a $k\times k$ matrix of the form
\be\label{Omega}
\Omega=\left(\begin{array}{cc} {\bf 1}_{\kp\times\kp} & 0 \\
                               0  & -{\bf 1}_{\km\times\km} \end{array}\right).
\ee

In the bulk of the D3-brane worldvolume the supersymmetry is still $N=4$, and the gauge group is
$U(k)$. Thus it is useful to think of the orbifold projection as imposing certain
boundary conditions on the $N=4$ $U(k)$ gauge theory living on the half-plane $x^6>0$.
From the point of view of $N=2$ SUSY the fields of this theory fall into a vector multiplet
of $U(k)$ and a hypermultiplet in the adjoint of $U(k)$. $\phi^\al$ and $\phi^i$ are the scalars
in the vector multiplet and hypermultiplet, respectively. In view of Eqs.~(\ref{proj},\ref{Omega})
it is convenient to represent the fields as block matrices of the form
$$ \left(\begin{array}{cc} A_{\kp\times\kp} & B_{\kp\times\km} \\
                            C_{\km\times\kp} & D_{\km\times\km} 
                            \end{array} \right) $$
We will refer to $A,D$ and $B,C$ as diagonal and off-diagonal components of a field, respectively.
Eq.~(\ref{proj}) implies that the diagonal components of the scalars in the vector multiplet
satisfy Neumann boundary conditions at $x^6=0$, while their off-diagonal components satisfy
Dirichlet boundary conditions. For the scalars in the hypermultiplet the role of Dirichlet and
Neumann conditions is reversed. By virtue of $N=2$ SUSY this completely determines the boundary
conditions for all fields at $x^6=0$.

Armed with this knowledge we now consider some more complicated brane configurations.
In order to obtain a setup which yields a three-dimen\-sional gauge theory at low energies, one
must terminate D3-branes at some object located at $x^6=L>0$. One possibility is to terminate them at
another orbifold 5-plane, so that the background becomes becomes ${\bf R}^6\times 
({\bf R}^3\times {\bf S}^1)/\cI$. Now every D3-brane is charged with respect to two $SO(2)$s. 
We will refer to the ``old'' $SO(2)$ as $SO(2)_1$, and call the new one $SO(2)_2$. Since $SO(2)_1$ and
$SO(2)_2$ charges can be chosen idependently, a variety of possibilities arise. We will consider 
only two special cases. 

In the first case, all D3-branes have the same charge with respect to 
$SO(2)_2$. This means that at the second orbifold plane all scalars in the vector multiplet 
have Neumann boundary conditions,
while all scalars in the hypermultiplet have Dirichlet boundary conditions. At energies much
smaller than $1/L$ only the modes which have Neumann boundary conditions on both ends survive.
Thus in this case the low-energy theory has only an $N=2$ vector multiplet of
$U(\kp)\times U(\km)$. This theory has only a Coulomb branch. The flat directions correspond to 
the motion of D3-branes along the orbifold plane. At a generic point of the moduli space, when all 
D3-branes are far apart, the gauge group is broken down to $U(1)^k$. When we bring all D3-branes 
together, it gets enhanced to $U(\kp)\times U(\km)$.

In the second case, we choose $SO(2)_2$ charges to be the same as $SO(1)_1$ charges.
Similar reasoning shows that the low-energy theory contains a vector multiplet of
$G=U(\kp)\times U(\km)$ and a pair of hypermultiplets in the $(\kp,\ov{\km})$ representation of
$G$. The theory has both Higgs and Coulomb branches. The Higgs branch corresponds to the
possibility that D3-branes of opposite $SO(2)$ charges coalesce, detach from the
orbifold planes, and move off in the $x^7,x^8,x^9$ directions. If $\kp\neq\km$, complete
Higgsing is impossible, so the Higgs branch is really a mixed Higgs-Coulomb branch. When we
bring all D3-branes together, $U(\kp)\times U(\km)$ is restored, and hypermultiplets become
massless. 

One can also terminate D3-branes on an NS5-brane parallel to the orbifold plane.
It follows from the analysis of Ref.~\cite{HW}
that the boundary conditions at the NS5-brane remove the hypermultiplet from the low-energy
spectrum, so the resulting low-energy theory is the same as in the first case above. 

\section{Mirror symmetry for $Sp(k)$ gauge theories in $d=3$}

Consider the following brane configuration: two D3-branes stretched between two orbifold 
planes located at $x^6=0$ and $x^6=L$, and $n-2$ NS5-branes parallel to the orbifold plane.
Let the positions of the NS5-branes in the $x^7,x^8,x^9$ be the same as those of 
the orbifold planes. We must also specify the $SO(2)$ charges of the D3-branes. 
We choose the two D3-branes to have opposite charges with respect to both $SO(2)_1$ and $SO(2)_2$.
The resulting configuration is shown in Figure 1a. Note that D3-branes
can break at the NS5-branes.
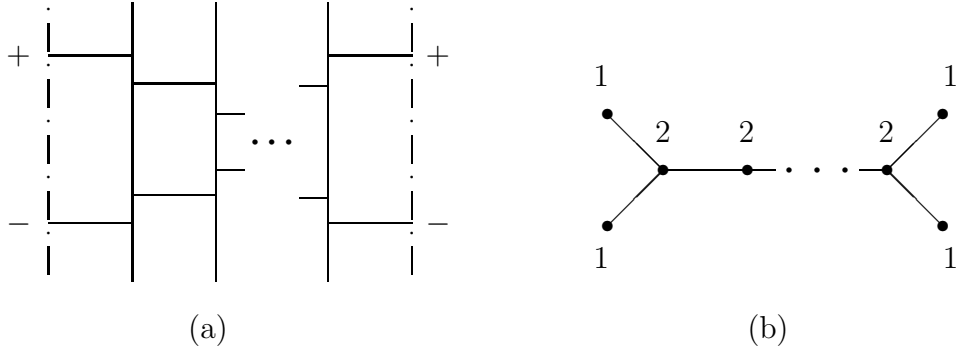
\begin{figure}

\setlength{\unitlength}{0.9em}
\begin{center}
\begin{picture}(34,13)

\savebox{\orbns}(0.5,10)[l]{\multiput(0,0)(0,2){5}{\line(0,1){1}}
\multiput(0,1.5)(0,2){5}{\circle*{.1}}}

\savebox{\ns}(0.2,10)[l]{\line(0,1){10}}

\savebox{\dbrane}(3,0.2){\line(1,0){3}}

\savebox{\dnquiver}(12,6){
\put(0,0){\line(1,1){2}}\put(0,4){\line(1,-1){2}}
\put(12,0){\line(-1,1){2}}\put(12,4){\line(-1,-1){2}}
\put(2,2){\line(1,0){4}}\put(10,2){\line(-1,0){1}}
\put(2,2){\circle*{.4}}\put(5,2){\circle*{.4}}
\multiput(6.5,2)(1,0){3}{\circle*{.2}}\put(10,2){\circle*{.4}}
\put(0,4){\circle*{.4}}\put(0,0){\circle*{.4}}
\put(12,4){\circle*{.4}}\put(12,0){\circle*{.4}}}

\put(2,2){\usebox{\orbns}}\put(5,2){\usebox{\ns}}\put(8,2){\usebox{\ns}}
\put(12,2){\usebox{\ns}}\put(15,2){\usebox{\orbns}}
\put(2,10){\usebox{\dbrane}}\put(2,4){\usebox{\dbrane}}\put(5,5){\usebox{\dbrane}}
\put(5,9){\usebox{\dbrane}}\put(12,4){\usebox{\dbrane}}\put(12,10){\usebox{\dbrane}}
\put(8,6){\line(1,0){1}}\put(8,8){\line(1,0){1}}
\multiput(9.4,7)(0.6,0){3}{\circle*{.2}}
\put(11,9){\line(1,0){1}}\put(11,5){\line(1,0){1}}
\put(0.5,9.8){$+$}\put(0.5,3.8){$-$}\put(15.5,9.8){$+$}\put(15.5,3.8){$-$}
\put(7,0){(a)}

\put(16,3){\usebox{\dnquiver}}
\put(21.5,9){1}\put(21.5,2.5){1}\put(23.7,7){2}\put(26.7,7){2}
\put(31.7,7){2}\put(34,9){1}\put(34,2.5){1}
\put(27,0){(b)}

\end{picture}\end{center}
\caption{(a) The dash-dotted lines are orbifold 5-planes. The solid vertical lines are NS5-branes.
The solid horizontal lines are D3-branes. The horizontal
direction corresponds to $x^6$, while the vertical direction 
corresponds to $x^3,\,x^4,$ and $x^5$ collectively. (b) $D_n$ quiver diagram encoding the gauge
group and matter content of the low-energy theory on the branes. Each node corresponds to
a unitary group, and the labels on them specify their rank. The solid lines connecting the nodes
correspond to bifundamental hypermultiplets.}
\end{figure}
From the analysis of the previous section it follows that the low-energy theory on D3-branes
is an $N=4$ gauge theory in $d=3$ with gauge group $U(1)^4\times U(2)^{n-3}$. The hypermultiplets
arise only from strings stretched across NS5-branes, as in Ref.~\cite{HW}. The gauge group
and matter content can be summarized by a quiver diagram~\cite{DouglasMoore}, which is nothing
but a $D_n$ affine Dynkin digram (Figure 1b). The nodes correspond to gauge group
factors, and the Dynkin labels tell us the rank of the unitary group sitting at a
particular node. It was argued in Ref.~\cite{IS} that this theory is mirror-symmetric
to an $SU(2)$ gauge theory with $n$ flavors. Now we can easily derive this statement
by applying S-duality to the above brane configuration. As discussed in the introduction, the 
S-dual of the orbifold plane is an $O5^-$ plane with a D5-brane on top of it. NS5-branes are dual
to D5-branes, and D3-branes are self-dual. We also have to figure out what the $SO(2)$ charges
of D3-branes  turn into under S-duality. To this end it is convenient to work on the double 
cover of ${\bf R}^6\times ({\bf R}^3\times {\bf S}^1)/\cI$, and to move the D5-branes slightly 
away from the $O5^-$ planes. The resulting S-dual configuration is shown in Figure 2. Note that
the $SO(2)$ charge being positive or negative corresponds to D3-branes ending on the D5-brane
or its mirror image. Now it is easy to see that the brane configuration in Figure 2 yields
at low energies an $SU(2)$ gauge theory with $n$ flavors of fundamentals.
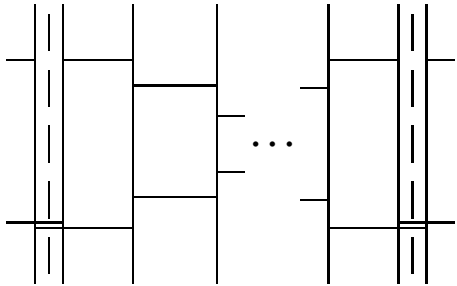
\begin{figure}

\setlength{\unitlength}{0.9em}
\begin{center}
\begin{picture}(17,13)

\savebox{\orb}(0.2,10)[l]{\multiput(0,0)(0,2){5}{\line(0,1){1.3}}}
\savebox{\ns}(0.2,10)[l]{\line(0,1){10}}
\savebox{\dbrane}(3,0.2){\line(1,0){3}}

\put(2,2){\usebox{\orb}}\put(1.5,2){\usebox{\ns}}\put(2.5,2){\usebox{\ns}}
\put(5,2){\usebox{\ns}}\put(8,2){\usebox{\ns}}\put(12,2){\usebox{\ns}}
\put(15,2){\usebox{\orb}}\put(14.5,2){\usebox{\ns}}\put(15.5,2){\usebox{\ns}}
\put(2.5,10){\line(1,0){2.5}}\put(1.5,10){\line(-1,0){1}}
\put(1.5,4){\line(1,0){3.5}}\put(2.5,4.2){\line(-1,0){2}}
\put(5,5){\usebox{\dbrane}}\put(5,9){\usebox{\dbrane}}
\put(12,10){\line(1,0){2.5}}\put(15.5,10){\line(1,0){1}}
\put(12,4){\line(1,0){3.5}}\put(14.5,4.2){\line(1,0){2}}
\put(8,6){\line(1,0){1}}\put(8,8){\line(1,0){1}}
\multiput(9.4,7)(0.6,0){3}{\circle*{.2}}
\put(11,9){\line(1,0){1}}\put(11,5){\line(1,0){1}}

\end{picture}\end{center}
\caption{The S-dual of the configuration in Figure 1a. The dashed lines are
$O5^-$ planes, the solid vertical lines are D5-branes, the solid
horizontal lines are D3-branes. For clarity we moved D5-branes slightly 
away from the orientifold planes. We are working on the double cover of the
orientifold background, but only part of it is shown in the picture.
In particular, the mirror images of only two D5-branes are shown.}

\end{figure}

This argument can be generalized. It has been argued in Ref.~\cite{Berkeley} that
the mirror of $Sp(k)$ gauge theory with $n$ fundamentals and an antisymmetric tensor is
the theory defined by the quiver in Figure 3b. To derive this statement from 
string theory we replace each D3-brane in Figure 1a with 
$k$ D3-branes with identical $SO(2)$ charges. The resulting configuration is shown in Figure 3a.
Reasoning similar to that in the previous paragraph shows that the S-dual brane configuration
yields an $Sp(k)$ gauge theory with $n$ fundamentals and an antisymmetric tensor. Thus
the mirror symmetry for these two gauge theories is a consequence of S-duality of IIB
string theory. 
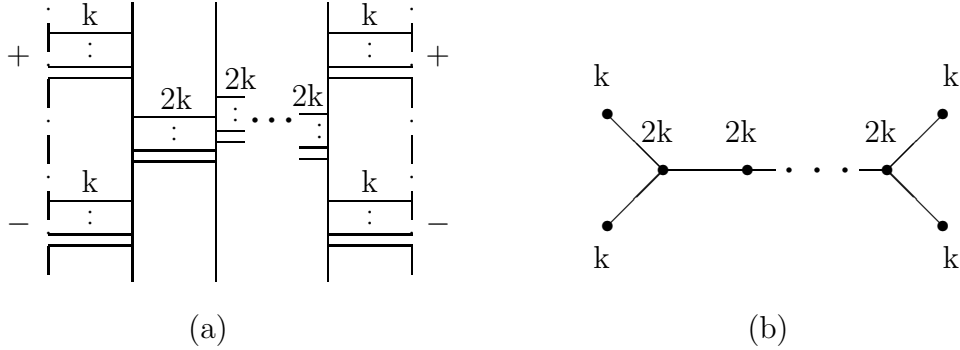
\begin{figure}

\setlength{\unitlength}{0.9em}
\begin{center}
\begin{picture}(34,13)

\savebox{\orbns}(0.5,10)[l]{\multiput(0,0)(0,2){5}{\line(0,1){1}}
\multiput(0,1.5)(0,2){5}{\circle*{.1}}}

\savebox{\ns}(0.2,10)[l]{\line(0,1){10}}

\savebox{\dbrane}(3,0.2){\line(1,0){3}}

\savebox{\bstack}(3,1.8){\put(0,0){\usebox{\dbrane}}
\put(0,0.4){\usebox{\dbrane}}\multiput(1.5,0.9)(0,0.4){2}{\circle*{.1}}
\put(0,1.6){\usebox{\dbrane}}}

\put(2,2){\usebox{\orbns}}\put(5,2){\usebox{\ns}}\put(8,2){\usebox{\ns}}
\put(12,2){\usebox{\ns}}\put(15,2){\usebox{\orbns}}
\put(2,9.2){\usebox{\bstack}}\put(3.2,11.2){k}
\put(2,3.2){\usebox{\bstack}}\put(3.2,5.2){k}
\put(5,6.2){\usebox{\bstack}}\put(6,8.2){2k}
\put(12,3.2){\usebox{\bstack}}\put(13.2,5.2){k}
\put(12,9.2){\usebox{\bstack}}\put(13.2,11.2){k}
\put(8,7){\line(1,0){1}}\put(8,7.4){\line(1,0){1}}
\multiput(8.7,7.8)(0,0.4){2}{\circle*{.1}}
\put(8,8.6){\line(1,0){1}}\put(8.3,8.9){2k}
\multiput(9.4,7.8)(0.6,0){3}{\circle*{.2}}
\put(11,6.4){\line(1,0){1}}\put(11,6.8){\line(1,0){1}}
\multiput(11.7,7.2)(0,0.4){2}{\circle*{.1}}
\put(11,8){\line(1,0){1}}\put(10.7,8.3){2k}
\put(0.5,9.8){$+$}\put(0.5,3.8){$-$}\put(15.5,9.8){$+$}\put(15.5,3.8){$-$}
\put(7,0){(a)}

\savebox{\dnquiver}(12,6){
\put(0,0){\line(1,1){2}}\put(0,4){\line(1,-1){2}}
\put(12,0){\line(-1,1){2}}\put(12,4){\line(-1,-1){2}}
\put(2,2){\line(1,0){4}}\put(10,2){\line(-1,0){1}}
\put(2,2){\circle*{.4}}\put(5,2){\circle*{.4}}
\multiput(6.5,2)(1,0){3}{\circle*{.2}}\put(10,2){\circle*{.4}}
\put(0,4){\circle*{.4}}\put(0,0){\circle*{.4}}
\put(12,4){\circle*{.4}}\put(12,0){\circle*{.4}}}

\put(16,3){\usebox{\dnquiver}}
\put(21.5,9){k}\put(21.5,2.5){k}\put(23.2,7){2k}\put(26.2,7){2k}
\put(31.2,7){2k}\put(34,9){k}\put(34,2.5){k}
\put(27,0){(b)}

\end{picture}\end{center}
\caption{(a) Brane construction of the mirror for
$Sp(k)$ gauge theory with an antisymmetric tensor and $n$ fundamentals.
The notation is the same as in Figure 1.
(b) Quiver diagram encoding the gauge group and matter content of the
brane configuration on the left.}

\end{figure}

An interesting special case of this mirror symmetry occurs for $n=2$ and $3$. 
Formally, the mirror of $Sp(k)$ theory with an antisymmetric tensor and $2$ or $3$ fundamentals
must be a $D_2$ or $D_3$ quiver theory, respectively.
However, $D_n$ quivers really make sense only for $n>3$. To figure out the answer for smaller $n$
one may start with a mirror pair for $n>3$ and then add a large mass term for some of the 
fundamentals of $Sp(k)$. In the mirror theory this corresponds to adding large FI
terms which Higgs the gauge group. From the point of view of brane configurations, this is
achieved by moving off some of the NS5 branes in the $x^7,x^8,x^9$ directions, so that
D3-branes reconnect. First consider integrating out all but three fundamental flavors, 
so that at low energies we have a theory with $n=3$. In the brane configuration in Figure 3a we move 
off all but one of the NS5 branes in the $x^7,x^8,x^9$ directions.
The resulting brane configuration is shown in Figure 4a. According to the rules
of the previous section, this theory has gauge group $U(k)^4$ and a matter
content summarized by the quiver in Figure 4b. All the hypermultiplets arise from
strings stretched across the NS5-brane. The quiver in Figure 4b is simply a perversely drawn 
affine $A_3$ quiver. Thus $D_3=A_3$, as one could have guessed.
To get the mirror for $n=2$ one moves off the last remaining NS5-brane, and gets the configuration
in Figure 5a. The gauge group is now $U(k)\times U(k)$, and the matter content is summarized by
the quiver in Figure 5b. It is the affine $A_1$ quiver, therefore $D_2=A_1$. Note that in this
case the hypermultiplets come from strings stretched``vertically'' between the two stacks of D3-branes.
These hypermultiplets make appearance only for $n=2$, because only in this case the boundary conditions
on both ends are Neumann.
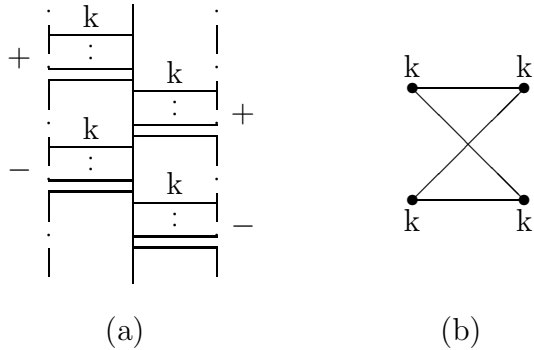
\begin{figure}

\setlength{\unitlength}{0.9em}
\begin{center}
\begin{picture}(20,13)

\savebox{\orbns}(0.5,10)[l]{\multiput(0,0)(0,2){5}{\line(0,1){1}}
\multiput(0,1.5)(0,2){5}{\circle*{.1}}}
\savebox{\ns}(0.2,10)[l]{\line(0,1){10}}
\savebox{\dbrane}(3,0.2){\line(1,0){3}}
\savebox{\bstack}(3,1.8){\put(0,0){\usebox{\dbrane}}
\put(0,0.4){\usebox{\dbrane}}\multiput(1.5,0.9)(0,0.4){2}{\circle*{.1}}
\put(0,1.6){\usebox{\dbrane}}}

\put(2,2){\usebox{\orbns}}
\put(5,2){\usebox{\ns}}
\put(8,2){\usebox{\orbns}}
\put(2,9.2){\usebox{\bstack}}\put(3.2,11.2){k}
\put(2,5.2){\usebox{\bstack}}\put(3.2,7.2){k}
\put(5,3.2){\usebox{\bstack}}\put(6.2,5.2){k}
\put(5,7.2){\usebox{\bstack}}\put(6.2,9.2){k}
\put(0.5,9.8){$+$}\put(0.5,5.8){$-$}\put(8.5,7.8){$+$}\put(8.5,3.8){$-$}
\put(4,0){(a)}

\put(15,9){\line(1,0){4}}\put(15,9){\line(1,-1){4}}
\put(15,5){\line(1,0){4}}\put(15,5){\line(1,1){4}}
\put(15,9){\circle*{.4}}\put(15,5){\circle*{.4}}
\put(19,9){\circle*{.4}}\put(19,5){\circle*{.4}}
\put(14.7,9.4){k}\put(18.7,9.4){k}
\put(14.7,3.9){k}\put(18.7,3.9){k}
\put(16,0){(b)}

\end{picture}\end{center}
\caption{(a) The brane construction of the mirror for $Sp(k)$ 
gauge theory with an antisymmetric tensor and $3$ fundamentals. 
The notation is the same as in Figure 1.
(b) Quiver diagram encoding the gauge group and matter content of the
brane configuration on the left.}

\end{figure}
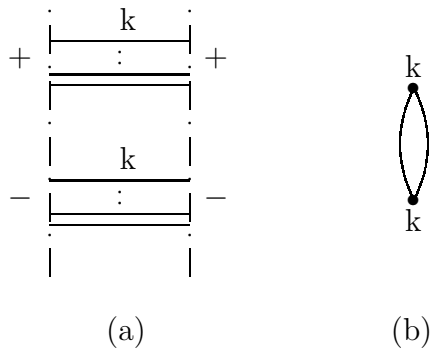
\begin{figure}

\setlength{\unitlength}{0.9em}
\begin{center}
\begin{picture}(16,13)

\savebox{\orbns}(0.5,10)[l]{\multiput(0,0)(0,2){5}{\line(0,1){1}}
\multiput(0,1.5)(0,2){5}{\circle*{.1}}}
\savebox{\dbrane}(5,0.2){\line(1,0){5}}
\savebox{\bstack}(5,1.8){\put(0,0){\usebox{\dbrane}}
\put(0,0.4){\usebox{\dbrane}}\multiput(2.5,0.9)(0,0.4){2}{\circle*{.1}}
\put(0,1.6){\usebox{\dbrane}}}

\put(2,2){\usebox{\orbns}}
\put(7,2){\usebox{\orbns}}
\put(2,9){\usebox{\bstack}}\put(4.5,11.1){k}
\put(2,4){\usebox{\bstack}}\put(4.5,6.1){k}
\put(0.5,9.8){$+$}\put(0.5,4.8){$-$}\put(7.5,9.8){$+$}\put(7.5,4.8){$-$}
\put(4,0){(a)}

\qbezier(15,9)(16,7)(15,5)\qbezier(15,9)(14,7)(15,5)
\put(15,9){\circle*{.4}}\put(15,5){\circle*{.4}}
\put(14.7,9.4){k}\put(14.7,3.9){k}
\put(14.2,0){(b)}

\end{picture}\end{center}
\caption{(a) The brane construction of the mirror for $Sp(k)$ 
gauge theory with an antisymmetric tensor and $2$ fundamentals. 
The notation is the same as in Figure 1.
(b) Quiver diagram encoding the gauge group and matter content of the
brane configuration on the left.}

\end{figure}

These results may seem puzzling. Indeed, a gauge theory described by an affine $A_n$
quiver has mass deformations~\cite{IS}. Its mirror must therefore have
FI deformations. But $Sp(k)$ theories do not have FI terms! We suggest the following resolution.
S-duality of IIB string theory seems to predict that superconformal theories which $Sp(k)$ 
theories in question flow to have deformations with the quantum numbers of 
FI terms. However these deformations need not have a simple description in 
terms of the gauge theory Lagrangian in the ultraviolet. Thus we are led to the conclusion that there
are ``hidden FI terms'' in $Sp(k)$ theories with an antisymmetric tensor
and $2$ or $3$ fundamentals. To count the number of hidden FI deformations we simply count the 
number of independent mass deformations of the mirror theories. In this way one can see 
that for $k=1$ there is only one hidden FI deformation, while for $k>1$ there are two. 

It is important to note that although $A_3$ and $A_1$ quiver theories have mass deformations,
they are not visible in our brane realization of these theories. We believe this is related 
to the fact that FI deformations in the mirror gauge theories are not visible classically. 

As a matter of fact, it is often the case that certain perturbations of a CFT do not have
a Lagrangian description in terms of the theory in the ultraviolet.
For example, the critical Ising model in d=2 is equivalent to a free Majorana fermion. 
Perturbation by magnetic field is described by the twist operator which is not local with 
respect to the fermion field. Therefore this perturbation does not have a Lagrangian description.
A more closely related example is afforded by the $N=4$ $d=3$ 
$U(k)$ gauge theory with an adjoint hypermultiplet and $n$ fundamentals.
The adjoint of $U(k)$ contains a singlet and an adjoint of $SU(k)$, hence this theory
has $n+1$ mass deformations: a mass term for the singlet, a mass term for the adjoint of $SU(k)$,
and $n-1$ mass terms for the fundamentals. The mirror theory~\cite{Berkeley} 
has only $n$ FI terms. A more detailed analysis reveals that it is the FI term corresponding to 
the mass difference of the singlet and the $SU(k)$ adjoint that is absent from the Lagrangian.
We conclude that the mirror theory has a hidden FI deformation.
Note again that the ``missing'' deformation, although present in field theory, is
not visible in the brane configuration of Ref.~\cite{Berkeley}.

The existence of FI deformations is equivalent to the existence of conserved $U(1)$ 
currents~\cite{five}. Thus our findings imply that $Sp(k)$ theories with $2$ or $3$ flavors
of fundamentals and an antisymmetric tensor have accidental $U(1)$ symmetries in the infrared.

$A_1$ and $A_3$ quiver theories are also mirror to $U(k)$ gauge theories with an adjoint and 
$2$ and $4$ flavors of fundamentals, respectively~\cite{Berkeley}. If we set $k=1$, we come to a
somewhat surprising conclusion that $U(1)$ theories with $2$ and $4$ electrons are equivalent
in the infrared to $SU(2)$ theories with $2$ and $3$ fundamentals, respectively. In the remainder
of this section we give some field theoretical arguments to support these equivalences. 

The moduli space of the $SU(2)$ theory 
with $3$ fundamentals consists of a Higgs branch of quaternionic dimension $3$ and a Coulomb 
branch of quaternionic dimension $1$ which meet at a single point~\cite{SW3}. 
The global $SO(6)$ flavor symmetry acts on the Higgs branch, but not on the Coulomb branch.
It is also the symmetry of the interacting conformal theory at the intersection of the two branches.
All these statement are also true for the $U(1)$ theory with $4$ electrons, essentially by
virtue of the isomorphism $SU(4)\simeq SO(6)$. The situation with the other equivalence is
a bit more involved. The Higgs branch of the $SU(2)$ theory with $2$ electrons consists of
two disconnected pieces of quaternionic dimension $1$~\cite{SW2}.
They intersect the Coulomb branch at two distinct points~\cite{SW3}. Each piece is isomorphic to
${\bf C}^2/{\bf Z}_2$, and the Coulomb branch has ${\bf Z}_2$ orbifold singularities
at the points where it meets the Higgs branch. The flavor group is $SO(4)\simeq SU(2)\times SU(2)$.
The two $SU(2)$ factors act separately on the two pieces of the Higgs branch. As a consequence,
the interacting conformal theory at the intersection of the Coulomb branch and one of the Higgs 
branches has only $SU(2)$ global symmetry, and not $SO(4)$. On the other hand, 
the $U(1)$ theory with $2$ electrons has a single Higgs branch isomorphic to ${\bf C}^2/{\bf Z}_2$.
The Coulomb branch also has a ${\bf Z}_2$ singularity at the intersection point~\cite{SW3}. The
global symmetry of the Higgs branch and the interacting conformal theory at the intersection is $SU(2)$.
Thus the moduli spaces and global symmetries of the $SU(2)$ theory with $2$ fundamentals and
$U(1)$ theory with $2$ electrons match in the neighborhoods of singular points. 
Note that in the strong coupling limit the two singular points on the Coulomb
branch of the $SU(2)$ theory are very far apart, so one is justified in considering only
the neighborhood of one of them. Thus field-theoretical analysis supports the proposed equivalences.

\section{$D_n$ quiver gauge theories in $d=4$}

In this section we use branes ending on orbifolds to construct a class of finite
$N=2$ $d=4$ gauge theories, along the lines of Ref.~\cite{Witten}. We explain the
relationship of these theories to instantons on ${\bf R}^2\times {\bf T}^2$ and 
give a simple derivation of their S-duality group. We then obtain an exact solution
of these theories using the method of Ref.~\cite{Kap}.

\subsection{$D_n$ quiver theories and $D_n$ instantons}
 
We start with the configuration in Figure 3a and T-dualize it along the $x^3$ direction. The
only effect of this T-duality is to turn D3-branes into D4-branes. The orbifold projection
remains unchanged, and IIB NS5-branes become IIA NS5-branes. The low-energy theory on the D4-branes
is now a $d=4$ $N=2$ theory, and its gauge group and matter content can be determined using 
the results of Section 2. One subtlety is the freezing of $U(1)$s due to the bending of
NS5-branes~\cite{Witten}. The rule of thumb is that all $U(1)$s that are not free will freeze out.
In the case at hand it implies that the gauge group is $SU(2k)^{n-3}\times SU(k)^4\times U(1)$. 
The $U(1)$ vector multiplet describes the motion of all D4-branes together along $x^4,x^5$.

The matter content of the low-energy theory is summarized by the $D_n$ quiver diagram (Figure 3b),
where each node now corresponds to a special unitary group.
We will call this theory a $D_n$ quiver theory, for short. Note that 
in Ref.~\cite{KMV} the $D_n$ quiver theory was defined without the $U(1)$ factor. Since 
the $U(1)$ vector multiplet is free, it plays almost no role in our discussion.

The $D_n$ quiver theory is finite and has interesting S-duality properties~\cite{KMV}. Namely,
its duality group (neglecting the $U(1)$) is the fundamental group of 
the moduli space of flat irreducible $SO(2n)$ connections on ${\bf T}^2$. More generally, it was shown in 
Ref.~\cite{KMV} that for a quiver to yield a finite $N=2$ gauge theory, it has to be one of the 
affine $A,D,E$ Dynkin diagrams. The duality group for an $A,D,E$ quiver theory is the fundamental group 
of ${\cal M}_{A,D,E}$, the moduli space of irreducible flat $A,D,E$ connections on ${\bf T}^2$. 
These results were obtained using geometric engineering. 

In the case of $A_n$ quiver theories alternative derivations 
are known~\cite{Witten,Kap}. In Ref.~\cite{Witten} $A_n$ quiver theories 
(there named ``elliptic models'') were constructed from branes in IIA string theory.
These models can be solved by lifting brane configurations to M-theory, and the
duality group turns out to be the fundamental group of ${\cal M}_{n+1}$, the moduli space
of an elliptic curve with $n+1$ marked points. One can easily see that this agrees with
Ref.~\cite{KMV}. These brane configurations can also be analyzed without resort to
M-theory~\cite{Kap}. To this end one compactifies the brane configuration on a circle
of finite radius and applies a sequence of T and S-dualities, thereby mapping the
Coulomb branch of the original theory to the Higgs branch of some mirror theory. The latter
does not receive quantum corrections and can be analyzed exactly. 
It is the last method that we will apply to $D_n$ quiver theories.  

Following Ref.~\cite{Kap}, we compactify the $x^3$ direction on a circle of radius $R$
and perform T-duality along it. This direction is parallel to all branes. It is easy
to see what the result is: the Neveu-Schwarz objects (NS5-branes and orbifold planes)
remain unchanged, while D4-branes become D3-branes. Next we perform S-duality, 
so that NS5-branes become D5-branes, and orbifold 5-planes become 
$O5^-$ planes with D5-branes on top of them. D3-branes are self-dual. The total number of
D5-branes is $n$, counting those which coincide with the $O5^-$ planes. Recall that 
in the original brane configuration we put NS5-branes on top of the fixed points
of the orbifold because we wanted to have a good CFT description of the background.
Now that the brane configuration contains only Ramond-Ramond objects,
we can move the D5-branes off the fixed points,
and still have a CFT description. This is important, since this deformation corresponds
to varying the relative gauge couplings of two $SU(k)$s at the ends of the quiver -- see
Figure 6.
\begin{figure}

\setlength{\unitlength}{0.9em}
\begin{center}
\begin{picture}(12,13)

\savebox{\ns}(0.2,10)[l]{\line(0,1){10}}

\savebox{\orb}(0.2,10)[l]{\multiput(0,0)(0,2){5}{\line(0,1){1.3}}}

\put(6,2){\usebox{\orb}}\put(5,2){\usebox{\ns}}
\put(7,2){\usebox{\ns}}\put(2,2){\usebox{\ns}}\put(10,2){\usebox{\ns}}
\linethickness{2pt}
\put(2,8){\line(-1,0){2}}\put(10,8){\line(1,0){2}}
\put(0.5,8.4){2k}\put(10.5,8.4){2k}
\put(5,10){\line(-1,0){3}}\put(7,10){\line(1,0){3}}
\put(3.2,10.3){k}\put(8.2,10.3){k}
\put(7,5){\line(-1,0){5}}\put(5,4.6){\line(1,0){5}}
\put(3.2,5.3){k}\put(8.2,4.9){k}

\end{picture}\end{center}
\caption{The dashed line is the IIA ``orbifold'' with
negative fivebrane charge. Solid vertical lines are NS5-branes.
Thick horizontal lines are stacks of D4-branes. We are working on the
double cover of the orbifold background, and only the neighborhood
of one fixed point is shown. When an NS5-brane and its mirror image
coincide with the fixed point, the gauge couplings of the two $SU(k)$s
at the end of the quiver are the same.}
\end{figure}
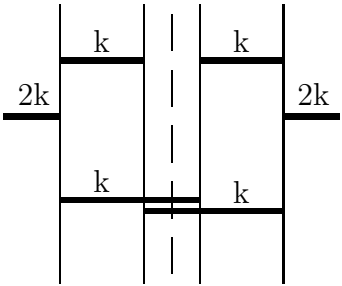

At this point there are
two ways to proceed. One possibility, extensively discussed in Ref.~\cite{Kap},
is to T-dualize back along $x^3$. This transformation turns D3-branes into D4-branes
wrapped around ${\bf T}^2$ parametrized by $x^3,x^6$. $O5^-$ planes turn into
$O4^-$ planes. Note that there are four $O4^-$ planes located at four points of ${\bf T}^2$, 
so D4-branes are effectively wrapped on ${\bf T}^2/{\bf Z}_2$. Finally, D5-branes 
turn into another set of D4-branes which we call D4$'$. D4$'$-branes
are located at points of the above-mentioned ${\bf T}^2$. Fundamental strings
stretched between D4-branes and D4$'$-branes give rise to hypermultiplets localized at points on
${\bf T}^2$. Thus the theory living on D4-branes is an impurity theory~\cite{Sethi,GS,KS}.
Its supersymmetric ground states correspond to solutions of certain first-order
differential equations on ${\bf T}^2$~\cite{KS}. These equations are Hitchin equations with
sources at the locations of the impurities. The sources are rank one matrices,
and their residues correspond to the masses of the bifundamentals in the
original theory~\cite{Kap}. The effect of the orientifolds is to impose a certain
projection on the connection and Higgs field appearing in the Hitchin equations~\cite{KS}.
The moduli space of Hitchin equations is the Coulomb branch of the $D_n$ quiver theory we started
from. In the next subsection we will use this to find the Seiberg-Witten curve for these
theories.

The second possibility is to T-dualize along $x^6$. $O5^-$ planes and D5-branes become an 
$O6^-$ plane and $n$ D6-branes wrapped on the ${\bf T}^2$ parametrized by $x^3,x^6$. This ${\bf T}^2$
is dual to the ${\bf T}^2$ on which the Hitchin equations live. As for D3-branes,
they become D2-branes parallel to D6-branes and localized in $x^3,x^6$. The gauge theory
on the D6-branes is an $SO(2n)$ theory with sixteen supercharges. D2-branes can be
viewed as BPS instantons of this theory. More precisely, they are instantons on
${\bf R}^2\times {\bf T}^2$, where ${\bf R}^2$ has coordinates $x^4,x^5$.  
Thus the Coulomb branch of the $D_n$ quiver theory on a circle is the same as the moduli space 
of $SO(2n)$ instantons on ${\bf R}^2\times {\bf T}^2$. 
To specify the moduli space of such instantons completely, one needs to fix their behaviour at 
infinity, where they become flat connections on ${\bf T}^2$~\cite{KS}. The moduli space of
flat connections being nontrivial, the Coulomb branch depends on the corresponding
moduli. These moduli are nothing but the gauge couplings and theta
angles of the $D_n$ quiver theory. To see this note that Hitchin equations
are the Nahm transform of instanton equations on ${\bf R}^2\times {\bf T}^2$, and the positions
of impurities encode the asymptotic behaviour of the instantons~\cite{KS}.
On the other hand, tracing the dualities one can easily see that the positions of impurities
on ${\bf T}^2/{\bf Z}_2$ are determined by the positions of NS5-branes
in the brane configuration in Figure 3a. Therefore these positions encode the gauge couplings of
the quiver theory.

To summarize, the moduli space of $2k$ $SO(2n)$ instantons on ${\bf R}^2\times {\bf T}^2$ 
asymptotic to a fixed flat connection is the Coulomb branch of the compactified
$D_n$ quiver theory, with complexified gauge couplings determined by the flat connection.
It follows that the S-duality group of the $D_n$ quiver theory is the fundamental group
of the moduli space of flat irreducible $SO(2n)$ connections on ${\bf T}^2$. This
agrees with the result of Ref.~\cite{KMV}. (The irreducibility condition arises 
because reducible flat connections correspond to some of the gauge couplings being infinite,
as one can easily convince oneself.) $A_n$ quiver theories are related to moduli spaces
of $U(n)$ instantons on ${\bf R}^2\times {\bf T}^2$ in a similar manner~\cite{Kap}.
Their duality group is therefore given by the $\pi_1$ of the moduli space of flat irreducible
$U(n)$ connections on ${\bf T}^2$, in agreement with Refs.~\cite{Witten,KMV}.

\subsection{Solution of the models}

As explained in the previous subsection, the moduli space of the $D_n$ quiver theory
compactified on a circle is identical to the moduli space of solutions of Hitchin equations
on a punctured ${\bf T}^2$ satisfying a certain projection condition. This projection
is the orientifold projection which acts on the coordinates of ${\bf T}^2$ by reflection.
Thus it has four fixed points.\footnote{Alternatively, the moduli space in question can be regarded
as the moduli space of {\it all} solutions of Hitchin equations on an orbifold Riemann
surface ${\bf T}^2/{\bf Z}_2$.} ${\bf T}^2$ with its complex structure is an elliptic curve
which we call $\Sigma$. Let $z$ be a ``flat'' holomorphic coordinate on $\Sigma$. The orientifold
projection acts by $z\ra -z$. The action on the connection and the Higgs field $\phi$ is of 
the form
\be\label{refl}
A_z(-z)=MA_z(z)^TM^{-1},\quad \phi(-z)=M\phi(z)^TM^{-1}
\ee
for some constant matrix $M$. To figure out $M$ one recalls that at the location of the
$O4^-$ plane the gauge group on D4-branes reduces from $U(2k)$ to $Sp(k)$. This means
that $M$ is a nondegenerate antisymmetric matrix. Without loss of generality we can
set it to
\be\label{M}
M=i\sigma_2\otimes {\bf 1}_{k\times k}.
\ee
Let us sketch an alternative way of deriving Eq.~(\ref{M}) from the properties of Nahm 
transform for $SO(2n)$ instantons. 
For simplicity, let us start with BPS-saturated $SO(2n)$ {\it monopoles}.
Any $SO(2n)$ monopole can be embedded into $U(2n)$, therefore Nahm data for $SO(2n)$ monopoles
are a subset of Nahm data for $U(2n)$ monopoles. This subset is easy to characterize~\cite{HM}:
it is a set of all $U(2n)$ Nahm data which satisfy the condition Eq.~(\ref{refl})
with $M$ as in Eq.~(\ref{M}), where $z$ is now a real parameter on which the Nahm
matrices depend. It is trivial to generalize this to instantons on ${\bf R}^3\times {
\bf S}^1$, and even to instantons on ${\bf R}^2\times {\bf T}^2$.

We thus have an implicit description of the hyperk\"ahler metric on the Coulomb branch
of the compactified $D_n$ quiver theory. In the decompactification limit (i.e.
when the radius of the compact circle is taken to infinity) only a distinguished
complex structure of this manifold is important~\cite{Kap}. This complex structure
is encoded in the spectral cover $C$ defined by the equation
\be\label{cover}
\det (v-\phi(z))=0.
\ee
This says that $C$ is a $2k$-fold cover of $\Sigma$. The importance of $C$ lies is in 
the fact that it is the Seiberg-Witten curve for the gauge theory we set out to 
solve~\cite{DW,Kap}. In the remainder of this paper we provide an explicit description
of $C$. As a first step, we rewrite Eq.~(\ref{cover}) in a more explicit form:
\be\label{curve}
v^{2k}+f_1(z) v^{2k-1}+\cdots+f_{2k}(z)=0.
\ee
As in Ref.~\cite{Kap}, Hitchin equations imply that $f_1,\ldots,f_{2k}$ are meromorphic
functions on $\Sigma$ with simple poles at the $2n$ punctures. The condition 
Eq.~(\ref{refl}) further implies that $f_1,\ldots,f_{2k}$ are even elliptic functions.
More concretely, let $\Sigma$ be given by
\be
y^2=(x-e_1)(x-e_2)(x-e_3).
\ee
An even elliptic function with $2n$ simple poles has the form
\be\nn
\sum_{i=1}^n\frac{a_i}{x-x_i}+b.
\ee
The poles are the solutions of ${\cal P}(z)=x_i,i=1,\ldots,n,$ where ${\cal P}(z)$ is the
Weierstrass elliptic function. We see that the space of even elliptic functions with $2n$
simple poles has dimension $n+1$. Therefore there are $2k(n+1)$ parameters in the
curve Eq.~(\ref{curve}). However, there is an additional constraint following
from Eqs.~(\ref{refl},\ref{M}): all the eigenvalues of $\phi(z)$ are doubly degenerate
at the four fixed points of the ${\bf Z}_2$ action $z\ra -z$. Therefore
all roots of Eq.~(\ref{curve}) are doubly degenerate for $x=e_1,e_2,e_3,\infty$.
This imposes $4k$ extra constraints on the coefficients of Eq.~(\ref{curve}),
thereby reducing the number of parameters to $2k(n-1)$. These parameters are interpreted
as the moduli of the Coulomb branch and the hypermultiplet masses. As in Ref.~\cite{Witten},
the asymptotic behaviour of $C$ is encoded in the residues of $f_1$, which are therefore
interpreted as the hypermultiplet masses. Thus we have $n$ hypermultiplet masses
and $2k(n-1)-n$ moduli. This agrees with the expected number of mass deformations
and moduli in the $SU(2k)^{n-3}\times SU(k)^4\times U(1)$ theory. 

A final remark concerns an extension of our results to $E_n$ quiver theories of
Ref.~\cite{KMV}. In view of the relation between $A_n$ and $D_n$ quiver theories
and $A_n$ and $D_n$ instantons, it is natural to expect that the exact solution
of $E_n$ quiver theories is described by a Hitchin system which is a Nahm transform of 
$E_n$ instantons on ${\bf R}^2\times{\bf T}^2$. Unfortunately, we do not know
what the Nahm transform looks like for exceptional groups. Neither is it clear what sort
of brane configuration could yield an exceptional quiver. We believe that a solution 
of the former problem could provide a key to the solution of the latter.

\renewcommand{\thesection}{Acknowledgements}
\section{}
I am grateful to N. Seiberg and A. Sen for illuminating discussions and for reading a 
draft of this paper. I also wish to thank O. Aharony for useful comments on
the manuscript.

\end{document}